# Internet of Things from Space: Transforming LTE Machine Type Communications for Non-terrestrial Networks

Talha Ahmed Khan, Xingqin Lin, Stefan Eriksson Löwenmark, Olof Liberg, Sebastian Euler, Jonas Sedin, Emre A. Yavuz, Hazhir Shokri-Razaghi, and Helka-Liina Määttänen

Ericsson

{talha.khan, xingqin.lin, stefan.g.eriksson, olof.liberg, sebastian.euler, jonas.sedin, emre.yavuz, hazhir.shokri.razaghi, helka-liina.maattanen}@ericsson.com

*Abstract*—Satellite communication is experiencing a new dawn thanks to low earth orbit mega constellations being deployed at an unprecedented speed. Fueled by the renewed interest in non-terrestrial networks (NTN), the Third Generation Partnership Project (3GPP) is preparing 5G NR, NB-IoT and LTE-M for NTN operation. This article is focused on LTE-M and the essential adaptations needed for supporting satellite communication. Specifically, the major challenges facing LTE-M NTN at the physical and higher layers are discussed and potential solutions are outlined.

## I. INTRODUCTION

Massive machine type communication (mMTC) is one of the three major use cases of the fifth generation (5G) system [1]. The mMTC use case is mainly about connecting a large number of low-complexity and low-cost devices in wide-area networks. Devices in the mMTC use case typically have a long battery life, feature a relatively low throughput, and require wide-area deep coverage. Example mMTC devices include meters, sensors, trackers, and wearables. Long-Term Evolution (LTE) for machine type communication (LTE-M) [2] and Narrowband Internet of Things (NB-IoT) [3] are optimized 5G access technologies for mMTC applications. They complement each other based on their unique characteristics: LTE-M can address a wide range of use cases including voice while NB-IoT uses narrow spectrum to cover ultra-low end mMTC applications [4]. It is predicted that there will be about 2.6 billion devices by 2026 connected via LTE-M and NB-IoT [5].

LTE-M and NB-IoT were first introduced in the 3rd Generation Partnership Project (3GPP) Release 13. Since then, they have been continuously evolved with new enhancements introduced in later releases. In the ongoing Release 17, there is a study item to explore the feasibility of adapting LTE-M and NB-IoT to support non-terrestrial networks (NTNs) [6]. NTNs will allow connectivity to be beamed from space (satellites) and the atmosphere (high altitude platform stations) to reach remote areas, thereby complementing terrestrial deployments. One example use case is logistics tracking in areas where terrestrial coverage is either not present or possible (e.g. maritime and air). The NTN work in 3GPP can be traced back to the 3GPP Release 15, during which a study on NTN scenarios and channel models for 5G New Radio (NR) was conducted [7]. In Release 16, 3GPP continued with a further study item on solutions for adapting NR to support NTNs [8]. In Release 17, besides the aforementioned study item on LTE-M and NB-IoT based NTNs, 3GPP is carrying out a work item to enable NR operation in NTNs [9]. An overview of the NR NTN work can be found in [10].

Machine type communications has attracted much interest in the past decade [11], but providing mMTC connectivity via NTNs is a relatively new area. The work [12] discussed the possible role of satellite systems in the mMTC use case and outlined some satellite mMTC system design trade-offs. The impact of the satellite channel characteristics on NB-IoT design was analyzed in [13]. The work [14] presented an interference study on satellite machine type communication for maritime Internet of Things (IoT). In [15], we reviewed the fundamentals of NB-IoT and NTN and explained how NB-IoT can be adapted to support satellite communication through a minimal set of essential adaptations.

The existing work has mainly focused on NB-IoT NTN and has not covered LTE-M NTN. The objective of this article is to address the gap by studying the feasibility of LTE-M NTN and identifying how LTE-M can be modified to support NTN with a minimal impact on the existing design. In particular, we present analysis and discussion on core topics including link budget calculation, solutions to Doppler, delay, and mobility challenges in NTN while striving for maintaining low device cost and high energy efficiency.

The remainder of this paper is organized as follows. Section II provides a brief technical introduction to LTE-M and NTN. The major challenges and potential solutions for LTE-M NTN are discussed in Section III for the physical layer, Section IV for the higher layers, and Section V for the architectural aspects. Section VI concludes the paper.

## II. TECHNICAL BACKGROUND

### A. LTE-M

LTE-M is part of the LTE standard and provides features to support the operation of low-cost, low-complexity and low-power mMTC devices. LTE-M networks can support the same system bandwidths as in LTE. LTE-M devices, however, are required to support a much smaller maximum bandwidth. It has two device categories (CAT): CAT-M1 and CAT-M2 for low-data rate and medium data-rate applications, respectively. CAT-M1 (resp. CAT-M2) devices support a maximum channel bandwidth of 1.4 MHz (resp. 5 MHz), and a maximum transport block size (TBS) of 2984 bits (or 6968 bits) on the uplink (UL) and 1000 bits (resp. 4008 bits) on the downlink (DL). LTE-M supports time division duplexing (TDD) and frequency division duplexing (FDD) with both half-duplex and full-duplex transmission modes.

LTE-M uses the same physical layer transmission schemes and numerologies as LTE. It uses orthogonal frequency-division multiplexing (OFDM) on DL and single-carrier frequency-division multiple access (SC-FDMA) on UL. The subcarrier spacing is 15 kHz for both DL and UL channels except for the physical random access channel (PRACH) which has a 1.25 kHz subcarrier spacing. A physical resource block (PRB) consists of 12 subcarriers spanning 180 kHz. This used to be the smallest possible resource that can be allocated to a user equipment (UE). Since Release 15, LTE-M also supports sub-PRB allocation on the physical uplink shared channel (PUSCH) which allows allocating only 30 kHz, 45 kHz or 90 kHz within a PRB.

LTE-M supports two modes to provide coverage enhancement (CE) based on physical layer repetitions: CE Mode A and CE Mode B. CE Mode A is used for moderate CE and allows up to 32 repetitions for the data channels. CE Mode B allows up to 2048 repetitions and can provide about 20 dB gain for devices requiring deep coverage. The network configures the device in one of the two modes based on the reported signal quality.

LTE-M inherits the radio protocol stack used in LTE. The user plane employs Packet Data Convergence Protocol (PDCP), Radio Link Control (RLC) and Medium Access Control (MAC) protocols to manage data transmission. In addition to these protocols, the control plane uses Radio Resource Configuration (RRC) protocol to provide signalling support.

LTE-M can be deployed within an LTE or an NR carrier. It can also connect to 5G core (5GC) since Release 16. It meets both the International Mobile Telecommunications (IMT)-2020 and the 3GPP 5G mMTC performance requirements in terms of connection density, data rate, maximum coupling loss, latency, and battery lifetime. Thus, it qualifies as a 5G technology.

*B. NTN*

In theory, there are endless possibilities for satellite orbits. In practice, however, three orbit types have proven popular and host the majority of all satellites: Low Earth Orbit (LEO), Medium Earth Orbit (MEO), and Geosynchronous Orbit (GEO).

LEO satellites typically orbit the Earth at altitudes between 500 km and 2000 km, corresponding to an orbital period of about 90 minutes to 2 hours. Assuming an altitude of 750 km, and a minimum elevation angle of about 10 degrees, each satellite is visible from an area on the surface of the Earth within a diameter of about 5000 km. Since these satellites move across the Earth surface with a velocity of 7.5 km/s, each satellite is visible to an observer on the ground for only about 10 minutes. With such a constellation, approximately 70 satellites are necessary to provide continuous global coverage.

At LEO altitudes, the drag produced by the upper atmosphere cannot be neglected. The resulting deceleration causes the orbits to decay, and to stay in orbit, LEO satellites continuously need to correct for this. Since fuel is limited, these correctional maneuvers reduce their lifetime to typically 5-10 years. The relative proximity to the surface of the Earth means that the round-trip delay is rather small (between 4 and 40 ms), and that the satellites have low power requirements and can be relatively small, which reduces the costs for construction and launch. LEO satellites have been used for communications since the 1990's, with the prime examples being the Iridium and Globalstar constellations. Recently, technological advances and significant reductions of the satellite launch costs have fueled proposals for new LEO constellations consisting of several thousands of satellites, e.g. by SpaceX (Starlink) and Amazon (Project Kuiper). These "massive LEO" constellations will multiply tenfold the number of satellites orbiting the Earth. Starlink has launched more than 1000 satellites to date, which means that they constitute about half of all active satellites.

GEO satellites have an altitude of 35,786 km, resulting in an orbital period of 24 hours which is synchronous with the Earth's rotation. A special case is an orbit with an inclination angle of 0° relative to the equatorial plane. For an observer on the ground, such a satellite appears to be at a fixed position in the sky. Therefore, such an orbit is called geostationary. If the inclination is different from 0°, the satellite moves north and south during the course of a day, forming a figure-8 pattern on the sky. Due to its high altitude, a single GEO satellite can cover a large fraction of the Earth, and only 3 satellites are enough to provide global coverage, except for the polar regions. The high GEO altitude, however, also creates challenges: To close the link, a high transmit power and satellite antennas with very high gain are needed. This increases the size and weight of the satellite and thus the costs of production and launch. For example, the TerreStar-1 satellite has a mass of around 7 t and features a foldable 18 m reflector antenna. Furthermore, the high altitude also causes large round-trip delays of more than 500 ms. GEO satellites have been used for decades, especially for broadcasting TV and radio (e.g., Eutelsat) by exploiting the possibility to install fixed antennas on buildings, but also for communications (e.g., Inmarsat Global Xpress).

There are different system architecture options to integrate a cellular network node into an NTN. With a bent-pipe or transparent architecture, the network node (i.e., base station) resides on the ground and is connected to the satellite via the satellite gateway. The satellite payload essentially acts as a relay between the base station and the user. In the regenerative architecture, the network functionality is hosted in the satellite payload itself. The regenerative architecture offers much lower latency than a bent-pipe architecture. The bent-pipe architecture, however, can expedite commercialization as existing satellite constellations can be adapted to support LTE-M operation.

Table I shows round-trip delays and Doppler shifts for different satellite orbits, assuming transparent architecture and a minimum elevation angle of 10 degrees. Note that for GEO there might be some residual Doppler shift due to satellite maneuvers for orbital corrections.

Table I Delay and Doppler shift values for various satellite orbits.

|  | LEO |  | GEO |
|---|---|---|---|
| Orbit altitude | 600 km | 1200 km | 35,786 km |
| Distance at min. elevation | 1932 km | 3131 km | 40,581 km |
| Delay at max. elevation | 8.0 ms | 16.0 ms | 477 ms |
| Delay at min. elevation | 25.8 ms | 41.8 ms | 541 ms |
| Max. Doppler shift | 24 ppm | 21 ppm | <1 ppm |
| Max. Doppler shift variation | 0.27 ppm/s | 0.13 ppm/s | ~0 ppm/s |

Satellites can generate one or more beams to connect to the users as well as the gateway. The satellite-gateway link is known as the feeder link while the satellite-user link as the service link. For the case of Earth fixed beams, the satellite uses a steerable antenna to direct its beam to a certain fixed region on the Earth surface until it goes out of view. For the case of moving beams, the satellite uses a fixed antenna and the beams constantly move with the satellite.

### III. LTE-M NTN: Physical layer aspects

The existing LTE-M technology faces many technical issues before it can be used for NTNs. For an expedited launch of LTE-M NTN, we need only focus on the essential adaptations needed to support NTN operation. In this section, we identify the key physical layer challenges facing LTE-M NTN and outline some potential solutions.

*A. Frequency bands*

LTE-M supports LTE FDD and TDD frequency bands ranging from 450 to 2690 MHz. Due to large round trip times (RTTs), it is challenging to support TDD in NTN. Therefore, LTE-M FDD bands should be prioritized. Moreover, to limit the impact on the current design, LTE-M NTN should only support the existing (sub 6 GHz) LTE-M frequency range

*B. Initial UL synchronization*

LTE-M was initially designed for terrestrial cellular networks which typically have a much smaller RTT and Doppler frequency shift than the differential RTT and differential Doppler frequency shift in an NTN. This poses several design challenges for initial timing acquisition during random access.

Firstly, the LTE-M PRACH was designed to estimate the UE timing assuming that the cyclic prefix (CP) duration exceeds the RTT. Larger RTTs can still be estimated but with a more advanced PRACH receiver implementation which uses a wider receive window and tests multiple time-of-arrival hypotheses. The receiver complexity scales with the maximum differential RTT that needs to be supported in the cell, which can be substantially large especially in a GEO NTN.

Secondly, the LTE-M PRACH design can only cope with a limited range of carrier frequency offset (CFO), e.g., up to half the subcarrier spacing when using preambles in unrestricted sets. As the maximum UL differential Doppler shift in a LEO NTN cell can be significantly larger, the PRACH receiver cannot estimate the uplink timing. A simple solution is to rely on a global navigation satellite system (GNSS)-equipped device in conjunction with the periodic broadcast of NTN satellite ephemeris information in the cell. The UE can obtain its position leveraging GNSS and the serving satellite's position from the ephemeris information. It can then calculate the UE-specific timing and frequency pre-compensation values to use for PRACH transmission. This means that the PRACH receiver needs to deal with only a residual timing error (smaller than the CP duration) amid a tolerable CFO. The UE further refines its timing based on the timing advance (TA) value in the random access response received from the network. This obviates the need of any changes to the existing PRACH design.

*C. TA maintenance*

The large timing and frequency drifts in an NTN make it challenging to maintain UL timing and frequency synchronization once a connection is established. In a terrestrial LTE-M cell, the network sends the TA maintenance command to a connected-mode UE. Using the same mechanism for maintaining timing synchronization in an NTN would result in a prohibitively large signalling overhead. For example, the timing drift can be as large as 40 us/s. A possible solution is to allow the UE to autonomously adjust its timing and frequency synchronization by leveraging GNSS and ephemeris information. This avoids the need of excessive control signalling from the network.

*D. Link Budget*

We present link budget results for GEO and LEO satellites for a representative set of simulation parameters (Set 1) considered in 3GPP [8]. We determine the receiver SNR as a function of the signal bandwidth (BW), the transmitter equivalent isotropically radiated power (EIRP), the receiver antenna gain-to-noise-temperature (G/T) ratio, the polarization loss (PL) between transmitter and receiver antennas, the free space path loss (FSPL), scintillation loss (SL), atmospheric loss (AL), shadow fading (SF) margin and Boltzmann's constant k, using

$$SNR = EIRP + \frac{G}{T} - 10\log_{10}(k) - FSPL - SF - SL - AL - PL - 10\log_{10}(BW) \quad (1).$$

In Table 2, based on the parameters considered in [8], we estimate the LTE-M SNR with sub-PRB transmission for a GEO NTN to be -3.04 dB in the DL and -3.19 dB in the UL. We note that both the physical downlink shared channel (PDSCH) and the PUSCH will require only a few repetitions (CE Mode A) for a 10% block error rate (BLER) thanks to the high SNR. For a LEO NTN, the SNR is 3.6 dB and 10.6 dB on DL and UL, respectively. A similar exercise will reveal CE mode A to be sufficient for the considered LEO NTN scenario.

Table 2 LTE-M link budgets for GEO and LEO NTNs for an S-band carrier frequency of 2 GHz.

|  | GEO | | LEO | |
|---|---|---|---|---|
|  | DL | UL | DL | UL |
| EIRP | 59.3 dBW | -7 dBW | 34.3 dBW | -7 dBW |
| G/T | -31.6 dB/K | 19.0 dB/K | -31.6 dB/K | 1.1 dB/K |
| BW | 1080 kHz | 30 kHz | 1080 kHz | 30 kHz |
| FSPL | 190.63 | 190.63 | 159.1 | 159.1 |
| AL | 0.190 | 0.190 | 0.1 | 0.1 |
| PL | 3 | 3 | 3 | 3 |
| SL | 2.2 | 2.2 | 2.2 | 2.2 |
| SF | 3 | 3 | 3 | 3 |
| SNR (dB) | -3.04 | -3.19 | 3.6 | 10.6 |

Certain NTN scenarios may suffer from a limited link budget on the UL due to limited UE transmit power and large path loss. For instance, with a 23 dBm EIRP and *full PRB* transmission (i.e., 180 kHz BW), the UL SNR will drop to -10.97 dB from the value reported in Table 2 for a GEO NTN. For the considered scenario, we can estimate the number of repetitions needed for data transfer. For UL data transfer, a physical layer data rate of about 7000 bps can be supported on PUSCH at a BLER of 10% and an SNR of -10.97 dB SNR. With a maximum TBS of 504 bits, this results in a transmission duration of about 72 ms which translates to 128 repetitions if full-PRB transmission is used. Therefore, CE mode B will be needed for this scenario. We note that a lower UE power class or a more pessimistic set of simulation parameters may make the link budget challenging even for a LEO NTN. Therefore, a potential solution is to leverage sub-PRB allocation to improve coverage: an UL allocation of 30 kHz, 45 kHz or 90 kHz can boost the UL SNR for GEO to -3.19 dB, -4.95 dB or -7.96 dB, respectively. Similarly, for LEO, the UL SNR can be improved to 10.6 dB, 8.8 dB or 5.8 dB, respectively. Moreover, sub-PRB (30 kHz) PUSCH transmission exhibits a peak-to-average-power ratio close to 0 dB, which makes it even more desirable in power-limited scenarios.

## IV. LTE-M NTN: HIGHER LAYER ASPECTS

In this section, we shed light on the major higher layer issues facing LTE-M NTN.

### A. Procedures and timers

Many LTE-M procedures are controlled through various timers in the MAC, RLC, PDCP and RRC layers. We list a few examples to show how timers are utilized in LTE-M:

- Random access procedure is controlled by timers that instruct the UE for how long it needs to monitor for responses from the network. Specifically, the *ra-ResponseWindowSize* is triggered after a UE sends the random access preamble, and the *mac-ContentionResolutionTimer* is started while the UE waits for the contention resolution message.
- Discontinuous Reception (DRX) mechanism allows the UE to monitor the DL control channel in an energy-efficient manner by waking up at preconfigured occasions rather than monitoring it continuously to receive a network message. It is also controlled by a set of timers.
- Scheduling Request (SR) message is sent by the UE on the UL control channel to request for resources from the network when it has data in its buffer to transmit. To control how often a UE can transmit a SR, *sr-ProhibitTimer* timer is used. It has a maximum value of 7 SR periods.
- RLC Automatic Repeat reQuest (ARQ) retransmission procedure also uses several timers. For instance, the *t-Reordering* timer is triggered when packets arrive at the receiver out-of-order due to packet loss at the MAC layer. Upon its expiry, the missing packet is considered lost and the receiver may send a status report to the transmitter to request retransmission. It has a maximum value of 1600 ms and a second-largest value of 200 ms.

Many of the LTE-M timers are already provisioned generously to support long transmission times and can cope with the large RTTs seen in an NTN. Some of these timers, however, may need to be adapted by either extending the timers or delaying their onset by an amount given by the RTT. We have identified two timers whose value ranges need to be extended: *sr-ProhibitTimer* and *t-reordering*. In addition, the *ra-ResponseWindowSize* and the *mac-ContentionResolutionTimer* should have their start offset by the RTT, similar to what was introduced in 5G NR NTN.

### B. Idle mode mobility

The existing LTE-M cell selection/reselection procedures are adopted as a baseline for LTE-M NTN. Similar to NB-IoT, LTE-M has a lower resolution in terms of Reference Signal Received Power (RSRP) and Reference Signal Received Quality (RSRQ) measurements due to lower complexity and a sparser measurement behavior. Due to the propagation characteristics of the satellite channel, RSRP/RSRQ values of neighboring cells are quite similar. The UE may have difficulties in determining the best cell to camp on if it solely relies on RSRP/RSRQ measurements. This issue is also relevant for neighbor cell RSRP/RSRQ measurements considering the relaxation rules which may not work as intended in an NTN cell since e.g., the UE may not be able to determine whether it is close to the cell edge based on RSRP/RSRQ measurements. A possible remedy is to let the UE use its location information and provide further assistance information such as the time a region is served, TA value to the target cell or elevation angles of the source and/or target cell to assist cell (re)selection [8], provided that such assistance information to be introduced has sufficient precision to work as expected for LTE-M.

Unlike GEO, the area served by a LEO satellite may change with the satellite motion. For example, for LEO with Earth fixed cells, the satellite serving a certain area will change as one satellite moves beyond horizon and a new satellite appears to serve the area. In other words, the service link is switched from one satellite to another. For NR NTN, it has been concluded that information on when a satellite cell will stop serving while another satellite cell will start serving a certain area can be broadcast by the network. The UEs may leverage this information in the cell (re)selection and idle mode measurement rules.

## C. Connected mode mobility

LTE-M supports connected mode mobility and it relies on the LTE/NR framework for measurements and handover. NTN scenario poses the following challenges:

- Increase in service interruption during handover due to large propagation delays.
- High handover rate due to rapid satellite movement.
- Reduced handover robustness due to small signal strength variation in regions of beam overlap.
- Propagation delay differences while measuring cells from different satellites.

In addition to the legacy handover mechanism, potential enhancements are to be considered also for conditional handover for both moving cell and Earth fixed cell scenarios. For example, the information mentioned about service link switch may be used as a conditional handover trigger along with the UE location with respect to the serving cell reference point.

## D. HARQ

LTE-M supports up to 10 HARQ processes in CE mode A and 2 HARQ processes in CE mode B. We note that the minimum time between initial transmission and received acknowledgment is determined by the RTT. This implies that only up to 10 DL data transmissions, one for each HARQ process, can be made every RTT. This will negatively impact the LTE-M peak data rates in an NTN.

To preserve the peak data rate, one idea is to aggressively increase the number of HARQ processes to accommodate the excessively large RTTs. The larger the RTT, the higher the number of HARQ processes needed. For instance, a few hundred HARQ processes are required to maintain the peak data rate in a GEO NTN. This is certainly not viable for a low-complexity LTE-M UE as it will adversely impact the UE receiver complexity and the required amount of buffer memory.

An alternative solution is to disable HARQ at the MAC layer and let RLC ARQ manage the retransmissions of failed transmissions. As RLC layer supports transmission of multiple blocks before feedback needs to be sent, it can lead to a significantly higher data rate than what can be supported with HARQ. Moreover, when HARQ is disabled, a lower target block error rate can buy more robustness.

The NR NTN work item considered both solutions. The number of HARQ processes were increased from 16 to 32 for LEO NTN systems with relatively small RTTs. Furthermore, network can disable selected number of HARQ processes by configuration. However, since throughput is not the key target for LTE-M, the needed enhancements will be studied with respect to the use cases and their requirements.

Let us consider the scenario where HARQ is disabled for PDSCH transmissions and RLC ARQ retransmissions are used. We first illustrate the gain of code combining enabled by HARQ in Figure 1. The residual BLER of a PDSCH after one, two and four HARQ transmissions is shown by the solid red, green and blue lines, respectively. In contrast, if only RLC retransmissions are used, the BLER after two and four transmissions are shown by the dashed green and blue lines, respectively. This dramatic difference is due to the use of code combining in the former case.

We note that code combining can also be achieved by blind repetitions (i.e., repetitions based on the transmitter's prediction of the required number of repetitions, not based on negative acknowledgements from the receiver) at the physical layer. As shown in Figure 2, the PDSCH BLER after four blind transmissions is almost the same as after four HARQ transmissions. Therefore, disabling HARQ does not necessarily mean a large performance loss, provided the link adaptation can select an appropriate modulation/coding scheme (MCS) and the number of blind repetitions. This is also illustrated in Figure 3. The solid blue line shows the average number of transmitted subframes needed to successfully receive one transport block of data when HARQ is used. The resource usage increases smoothly as the SNR is decreased. The dashed red line shows the corresponding resource usage with four blind transmissions on the physical layer when RLC ARQ is used to correct the residual errors. At high SNR, the resource usage is lower bounded by the number of blind transmissions, i.e., four. Around -3 to -4 dB SNR, the resource usage is only slightly worse than when HARQ is used. However, as the SNR is further reduced, the resource usage rapidly increases. It is, therefore, important to dynamically adapt the MCS and/or the number of blind repetitions to the SNR when HARQ feedback is disabled.

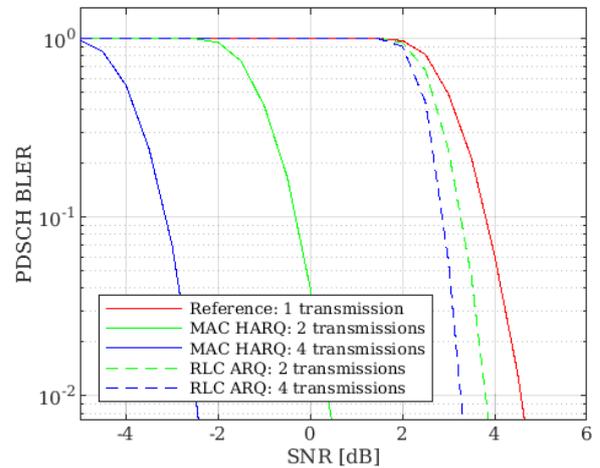

Figure 1 PDSCH HARQ versus ARQ retransmission BLER performance.

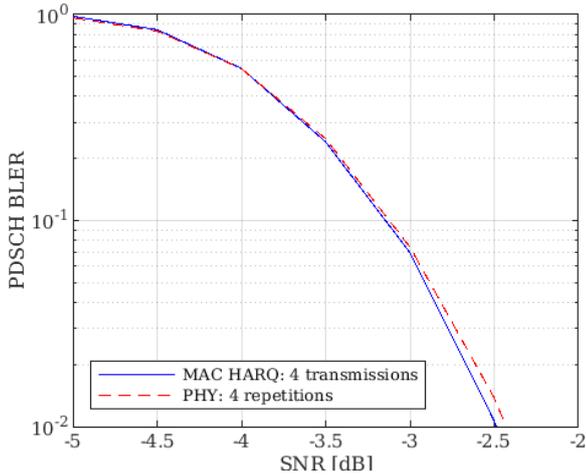

Figure 2 PDSCH HARQ retransmission versus blind repetition BLER performance.

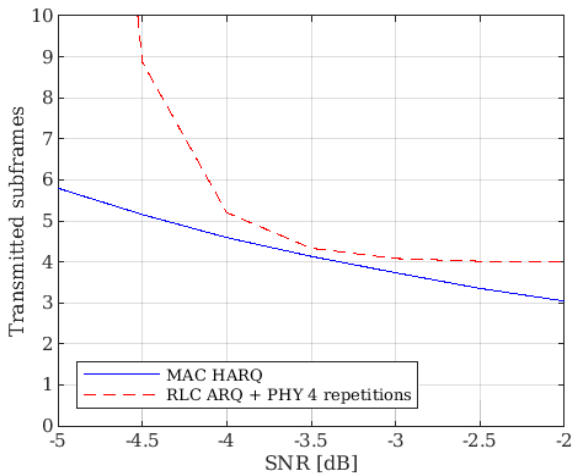

Figure 3 PDSCH HARQ retransmission versus blind repetition resource usage.

## V. LTE-M NTN: Architectural aspects

In this section, we discuss the architectural aspects of LTE-M NTN. LTE-M is served by the Evolved Packet Core (EPC) architecture since its inception. Since Release 16, it can also connect to the 5GC. EPC support should be prioritized for LTE-M over NTN as it is predominantly used in existing deployments. There are no essential changes needed for EPC to provide the core functionality for LTE-M NTN.

The ongoing Release 17 study on LTE-M NTN focusses on NTNs with transparent satellite payloads. This is to facilitate the integration of LTE-M to existing NTNs that support transparent satellite payloads. In the 3GPP terminology, the satellite node with a transparent payload is essentially an advanced repeater with the added capability to shift frequencies between the feeder link and the service link frequency bands. The existing 3GPP specifications need to be adapted to support such a sophisticated repeater node for LTE-M.

In Table 3, we summarize the main challenges for LTE-M NTN and their potential solutions.

*Table 3 Overview of key issues facing LTE-M NTN and potential solutions.*

| Issues | Solutions |
|---|---|
| UL synchronization amid large delay and Doppler shift | Use GNSS and ephemeris information to estimate and pre-compensate for RTT and Doppler shift before random access. |
| Large timing and frequency drift | UE performs autonomous timing and frequency adjustment. |
| Low UL SNR | Use sub-PRB allocation. |
| Higher layer procedures | Extend *sr-ProhibitTimer* and *t-reordering* timer ranges. Delay the onset of *ra-ResponseWindowSize* and the *mac-ContentionResolutionTimer* timers. |
| Limited peak data rate | Disable HARQ and rely on higher layer retransmissions. |
| Mobility | Use UE location along with assistance information. |
| NTN-specific signalling support | Use RRC signalling to send assistance information. |
| Transparent architecture | Support advanced repeater node with frequency conversion capability. |

## VI. Conclusions

In this article, we have provided a brief introduction to LTE-M in the context of NTN. We have discussed how LTE-M technology can be enhanced to address the unique challenges posed by NTN at the physical and higher layers. We have identified the essential changes needed to enable LTE-M operation in NTNs. At the physical layer, timing and frequency pre-compensation and autonomous adjustment of timing and frequency reference are needed to cope with the large delay and Doppler shifts. At the higher layers, a few timers require an extension in the value range or a delayed onset to account for the large delay. Moreover, the network may broadcast assistance information to facilitate mobility. Furthermore, HARQ mechanism can be disabled at the MAC layer and RLC retransmissions can be used for robustness, as depicted in the simulation results. With the proposed modifications, LTE-M is expected to support both LEO and GEO constellations.

### References

[1] X. Lin and N. Lee, "5G and Beyond: Fundamentals and Standards," Springer, 2021.

[2] A. Hoglund et al., "Overview of 3GPP Release 14 Further Enhanced MTC," *IEEE Communications Standards Magazine*, vol. 2, no. 2, pp. 84-89, June 2018.

[3] Y. E. Wang et al., "A Primer on 3GPP Narrowband Internet of Things," *IEEE Communications Magazine*, vol. 55, no. 3, pp. 117-123, March 2017.

[4] O. Liberg et al., "Cellular Internet of things: technologies, standards, and performance," Academic Press, 2017.

[5] Ericsson, "Ericsson mobility report," white paper, November 2020.

[6] RP-202689, "Study on NB-IoT/eMTC support for Non-terrestrial Network," 3GPP TSG RAN#90, December 2020.

[7] TR 38.811, "Study on New Radio (NR) to support non-terrestrial networks," 3GPP, V15.4.0, October 2020.

[8] TR 38.821, "Solutions for NR to support Non-terrestrial Networks (NTN)," 3GPP, V16.0.0, January 2020.

[9] RP-202909, "Solutions for NR to support non-terrestrial networks (NTN)," 3GPP TSG RAN #90e, December 2020.

[10] X. Lin et al., "5G new radio evolution meets satellite communications: Opportunities, challenges, and solutions," in *5G and Beyond: Fundamentals and Standards*, X. Lin and N. Lee, Eds. Springer, 2021.

[11] Z. Dawy, W. Saad, A. Ghosh, J. G. Andrews and E. Yaacoub, "Toward massive machine type cellular communications," *IEEE Wireless Communications*, vol. 24, no. 1, pp. 120-128, February 2017.

[12] S. Cioni, R. De Gaudenzi, O. Del Rio Herrero and N. Girault, "On the satellite role in the era of 5G massive machine type communications," *IEEE Network*, vol. 32, no. 5, pp. 54-61, September/October 2018.

[13] A. Guidotti et al., "Architectures and key technical challenges for 5G systems incorporating satellites," *IEEE Transactions on Vehicular Technology*, vol. 68, no. 3, pp. 2624-2639, March 2019.

[14] T. Xia, M. M. Wang and X. You, "Satellite machine-type communication for maritime Internet of things: An interference perspective," *IEEE Access*, vol. 7, pp. 76404-76415, May 2019.

[15] O. Liberg et al., "Narrowband Internet of things for non-terrestrial networks," *IEEE Communications Standards Magazine*, December 2020.